\documentclass{jpsj-suppl}
\usepackage{txfonts} 

\title{Properties of Odd-frequency Superconductivity in Antiferromagnetic Ordered State}

\author{Tomonori Harada, Yuki Fuseya and Kazumasa Miyake}

\inst{Divisin of Materials Physics, Department of Materials Engineering Science, Graduate School of Engineering Science, Osaka University, Toyonaka, Osaka 560-8531, Japan}

\email{miyake@mp.es.osaka-u.ac.jp}

\recdate{November 16, 2011}

\abst{\hspace{1em}
We investigate properties below $T_{\textrm{C}}$ of odd-frequency pairing which is realized by 
antiferromagnetic critical spin fluctuations or spin wave modes. It is shown that 
$\Delta(\epsilon_n)$ becomes maximum at finite 
$\epsilon_n$, and $\Delta(\pi T)$ becomes maximum at finite $T$. Implications of the present 
results to the experimental results of CeCu$_2$Si$_2$ and CeRhIn$_5$ are given.}

\kword{odd-frequency, gapless superconductivity, coexistence of antiferromagnetism and superconductivity, 
CeCu$_2$Si$_2$, CeRhIn$_5$}
\begin{document}
\maketitle
\section{Introduction}
In Ce-based heavy fermion compounds, it is known that CeCu$_2$Si$_2$ and CeRhIn$_5$, 
under pressure, exhibit 
two kinds of superconductivity (SC), gapless SC and line-node gap SC, from measurements of the 
nuclear spin-lattice relaxation rate 1/T$_1$~\cite{Y_Kawasaki,Mito,S_Kawasaki}.  
At low pressure side of the "critical pressure", 
antiferromagnetism (AF) and the gapless SC coexist below the superconducting transition temperature 
$T_{\textrm{C}}$. On the other hand, at high pressures, AF disappears and the line-node SC appears. 
This gapless SC is not due to impurity scatterings, because the clear line-node gap SC recovers with 
the same sample at pressures exceeding the critical pressure. Namely, the gapless SC seems to be 
an intrinsic and a novel SC state.

It was pointed out that the gapless nature can be understood as the odd-frequency 
\textit{p}-wave singlet pairing is occurring by critical spin fluctuations and antiferromagnetic spin 
waves~\cite{Fuseya}.  However, previous theories discussed only a behavior of $T_{\textrm{C}}$. 
Properties of SC below $T_{\textrm{C}}$ 
have not been understood yet. In this paper, we investigate these properties below $T_{\textrm{C}}$ 
by using the same spin fluctuation modes as used in Ref.\ \citen{Fuseya}.  In particular, 
we examine property of the frequency dependent gap function $\Delta(\epsilon_{n})$, $\epsilon_{n}$ 
being the fermionic Matsubara frequency, by solving the gap equation.  
It is shown that $\Delta(\pi T)$ takes maximum at finite temperature 
and decreases towards zero as temperature decreases. When the effect of spin fluctuations is not strong, 
the odd-frequency SC exhibits the reentrant behavior.  On the other hand, the odd-frequency SC remains 
even at zero temperature when the effect of critical spin fluctuation is strong enough at the criticality 
or in the ordered state of AF. 

\section{Theory}
We introduce the pairing interaction mediated by critical antiferromagnetic spin fluctuations as follows: 
\begin{equation}
V(\textbf{q},\omega_m)=g^2\chi(\textbf{q},\omega_m)=\frac{g^2N_\textrm{F}}{\eta+A\hat{\textbf{q}}^2+C|\omega_m|},
\label{e1}
\end{equation}
where $g$ is the coupling constant, $N_\textrm{F}$ the density of states at the Fermi level, and 
$\hat{\textbf{q}}^2\equiv4+2(\cos q_x +\cos q_y)$  in two dimensions. This type of pairing interaction 
was adopted by Monthoux and Lonzarich to discuss the strong coupling effect on the 
superconducticity induced by the critical AF fluctuations~\cite{Monthoux}.  The parameter $\eta$ 
in eq.\ (\ref{e1}) parameterizes a distance from the QCP.  We can treat coexistence phase by 
setting $\eta=0$.  

The pairing interaction can be decomposed as 

\begin{equation}
V_{\ell}(\epsilon_n-\epsilon_{n^\prime})=\sum_{\textbf{k}, \textbf{k}^\prime}\phi_{\ell}(\textbf{k})
V_{\ell}(\textbf{k}-\textbf{k}^\prime,\epsilon_n-\epsilon_{n^\prime})
\phi_{\ell}^*(\textbf{k}^\prime)\simeq 
v_{\ell}\ln\frac{\epsilon_\textrm{F}}{\sqrt{(\epsilon_n-\epsilon_{n^\prime})^2+\eta^2}},
\label{e2}
\end{equation}
where $\epsilon_{n}$ and $\epsilon_{n^\prime}$ are Matsubara frequencies.  
The interaction, eq.\ (\ref{e2}),  exhibits logarithmic divergence when 
$\epsilon_n-\epsilon_{n^\prime}\simeq0$ at the criticality $\eta=0$.  
Sharper the divergence is obtained by using smaller the value of the paramter $\eta$.
We introduce a cutoff $\epsilon_\textrm{F}$ and restrict Matsubara frequencies such that 
$(\epsilon_n-\epsilon_{n^\prime})_\textrm{max}\le\epsilon_\textrm{F}$.  
The coupling constant $v_{\ell}$ is a positive coefficient which is determined by the relation 
between the Fermi surface and AF ordering vector. $\ell$ takes even or odd integer indicating 
type of paring.  When AF ordering vector is comparable to a diameter of the Fermi surface 
without nesting tendency, $p$\/-wave ($\ell=1$) odd-frequency pairing is promoted against 
$d$-wave ($\ell=2$) even-parity one.    
In this paper, we suppose $v_{\rm o}$ is larger than $v_{\rm e}$, since we consider the situation 
that the odd-frequency is promoted by AF background~\cite{Kusunose}.

The gap equation (non linearized) is given as 

\begin{equation}
\Delta_{{\rm e},{\rm o}}(\epsilon_n)=-k_\textrm{B}T\sum_{n^\prime, k^\prime} 
V_{{\rm e},{\rm o}}(\epsilon_n-\epsilon_{n^\prime})
\frac{\Delta_{{\rm e},{\rm o}}(\epsilon_{n^\prime})}
{{\epsilon_{n^\prime}}^2+{\xi_{k^\prime}}^2+|{\Delta_{{\rm e},{\rm o}}(\epsilon_{n^\prime})}|^2},
\label{e3}
\end{equation}
where $\xi_{k^\prime}$ is quasiparticle energy measured from the chemical potential (Fermi level).  
We solved this gap equation selfconsistently.
\section{Result}
\subsection{Frequency dependence of gap function} 

The frequency dependence of gap function is shown in Fig.1 for 
$T/T_\textrm{C}=0.99,\hspace{1ex} 0.25,\hspace{1ex} 0.01$.  Near $T_\textrm{C}$, the relation 
$\Delta_{{\rm o}}(\epsilon_n)\propto1/\epsilon_n$ roughly holds. 
$\Delta_{{\rm o}}(\epsilon_n)$ takes maximum at the lowest frequency $\epsilon_0=\pi T$ as 
in the ordinary even-frequency SC.  
On the other hand, at low tmperature $T/T_\textrm{C}=0.01$, $\Delta_{o}(\epsilon_n)$ takes 
maximum at finite $\epsilon_{n}$, reflecting the odd frequency nature, while 
the relation $\Delta_{e}(\epsilon_n)\propto1/\epsilon_n$ roughly holds  
in the whole temperature region in the case of even-frequency SC.

\begin{figure}[tbh]
\begin{center}
\includegraphics[width=0.66\linewidth]{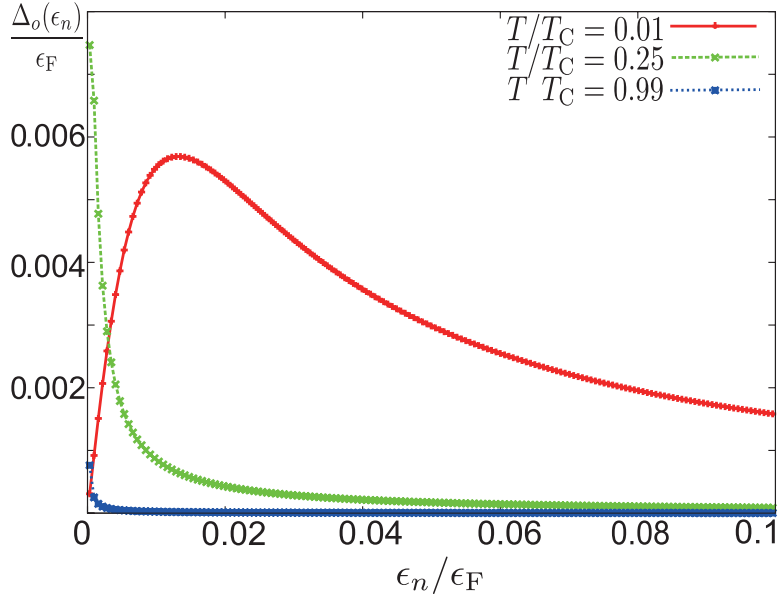}
\caption{Frequency dependences of gap function in the case of $\eta=0.005$ 
for a series of temperatures, $T/T_{\rm C}$=0.01, 0.25, and 0.99. }
\label{f1}
\end{center}
\end{figure}

\subsection{Temperature dependence of gap function}

The temperature dependence of gap function is shown in Fig.2 for 
$\eta=0.008,\hspace{1ex} 0.005,\hspace{1ex} 0.001$.  Solid lines represent $\Delta_{{\rm o}}(\pi T)$.  
Broken lines represent $\Delta_{{\rm o}\,\textrm{max}}(\epsilon_n)$. 
$\Delta_{{\rm o}\,\textrm{max}}(\epsilon_n)$ coincides with $\Delta_{{\rm o}}(\pi T)$ in high 
temperature region.  However, $\Delta_{{\rm o}\,\textrm{max}}(\epsilon_n)$ deviates from 
$\Delta_{{\rm o}}(\pi T)$ in low temperature region. This difference arises from reduction of gap function 
in low frequency region, as shown in Fig.1.

$\Delta_{{\rm o}}(\pi T)$ takes a maximum at around $T\simeq T_\textrm{C}/2$ and 
$\Delta_{{\rm o}}(\pi T)$ decreases as temperature decreases. This reduction is in contrast to the 
case of ordinary even-frequency gap, in which $\Delta_{{\rm e}}(\pi T)$ saturates at low 
temperatures. When $\eta=0.008$, $\Delta_{{\rm o}}(\pi T)$ vanishes at low temperature side, exhibiting 
reentrant behavior. Whereas for $\eta=0.005,\hspace{1ex} 0.001$ and $\Delta_{{\rm o}}(\pi T)$ is 
finite in the whole region of $T\le T_\textrm{C}$. Maximum at finite $T$ of $\Delta_{{\rm o}}(\pi T)$ 
is due to the maximum at finite $\epsilon_n$ of $\Delta_{o}(\epsilon_n)$ at low temperatures.

\begin{figure}[tbh]
\begin{center}
\includegraphics[width=0.8\linewidth]{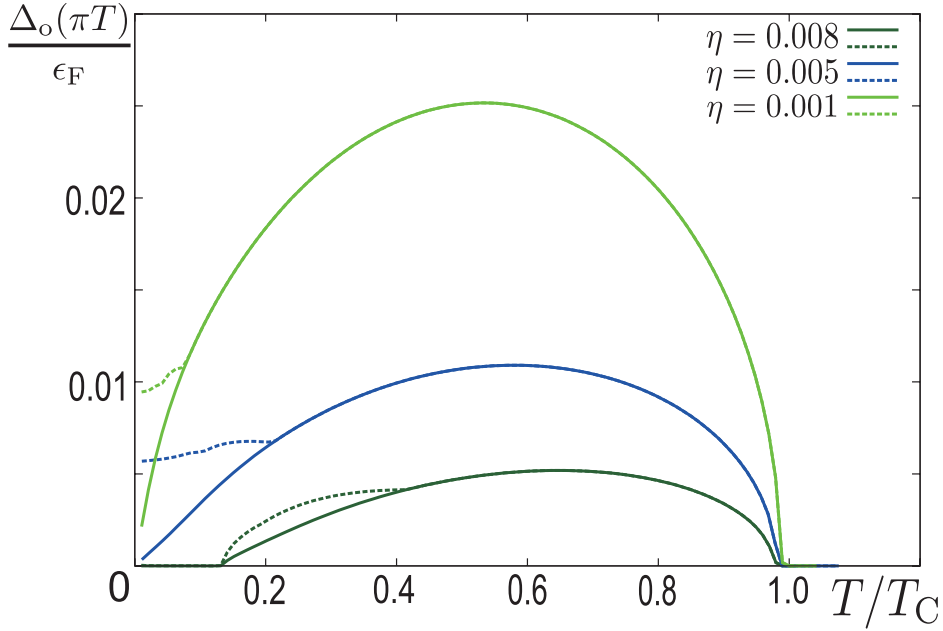}
\caption{Temperature dependence of gap function $\Delta_{{\rm o}}(\pi T)$.  
Broken lines represent $\Delta_{{\rm o}\,\textrm{max}}(\epsilon_n)$.}
\label{f2}
\end{center}
\end{figure}

\section{Conclusion}

We have solved gap equation for odd-frequency pairing realized by the antiferromagnetic 
critical spin fluctuations or spin wave modes of AF.  The gap function of odd-frequency SC takes 
maximum with respect to $\epsilon_n$ and $T$. 
Reentrant behavior occurs when $\eta$ is not extremely small. On the other hand, the odd-frequency SC 
is realized when $\eta$ approaches zero.  
Thus, the coexistence phase of SC and AF observed in CeCu$_2$Si$_2$ and CeRhIn$_5$ can be 
simulated by our model with zero $\eta$ limit. In this situation, $\Delta_{{\rm o}}(\pi T)$ is finite 
even at $T=0$.

At very low temperature region, however, odd-frequency can compete with even-frequency. 
If we consider this competition, transition from odd-frequency to even-frequency is possible 
at very low temperature.  This problem will be discussed elsewhere.

\end{document}